\documentclass[12pt]{article}

\usepackage{graphicx}

\usepackage{amssymb}
\usepackage{amsmath}

\title{Hydrodynamics of the Oscillating Wave Surge Converter in the open ocean}

\author{E. Renzi - F. Dias}
\author{School of Mathematical Sciences, University College Dublin, \\Belfield, Dublin 4, Ireland}
\date{Submitted to: EJMB/Fluids, 16/07/2012}

\begin{document}
\maketitle
\section*{Abstract}
A potential flow model is derived for a large flap-type oscillating wave energy converter in the open ocean. Application of the Green's integral theorem in the fluid domain yields a hypersingular integral equation for the jump in potential across the flap. Solution is found via a series expansion in terms of the Chebyshev polynomials of the second kind and even order. Several relationships are then derived between the hydrodynamic parameters of the system. Comparison is made between the behaviour of the converter in the open ocean and in a channel. The degree of accuracy of wave tank experiments aiming at reproducing the performance of the device in the open ocean is quantified. Parametric analysis of the system is then undertaken. It is shown that increasing the flap width has the beneficial effect of broadening the bandwidth of the capture factor curve. This phenomenon can be exploited in random seas to achieve high levels of efficiency.





\section{Introduction}
\label{sec:intro}
Recent analysis \cite{WF12} has shown that the oscillating wave surge converter (OWSC) is an effective contender in the challenge to extract energy from water waves. In its simplest form, the OWSC is a bottom-hinged buoyant flap used to extract wave power in the nearshore environment, where water particles move essentially in surge.

Traditional research in wave energy has mostly focused on two-dimensional (2D) devices \cite[see for example]{EV76, SR80}, on three-dimensional floating bodies in deep water \cite{EV76} and on buoys of characteristic width much smaller than the incident wavelength, known as ``point absorbers'' (PA) \cite{BU77, FA02}. Regrettably, the potential of the OWSC has been somewhat disregarded in the past. Given the incident wave wavelength $\lambda$ and the flap width $w$, traditional theories are indeed applicable to the OWSC in the 2D approximation $w\gg\lambda$ and in the PA limit $w\ll \lambda$ \cite[see for example]{FA02, ME05}. However, there is still a gap of knowledge in the region $w=O(\lambda)$, which is the most likely for practical applications \cite{WF12}. Lately, Falnes and  Hals \cite{FH12} proposed to treat wave energy converters (WECs) for which $w=O(\lambda)$ as ``quasi-point absorbers'' (QPA). However, such definition is somehow ambiguous: a precise mathematical formulation is needed to determine the hydrodynamic behaviour of such WECs in detail. The mathematical models of Renzi and Dias \cite{RD12a, RD12b} partially achieve this goal. They describe the resonant behaviour of a single OWSC in a channel and of a periodic array of OWSCs. Within the framework of a potential flow theory, Renzi and Dias \cite{RD12a, RD12b} show that the OWSCs can attain high levels of capture factor, with peaks occurring at resonance of the transverse short-crested waves of the channel/array system. Despite the latest advancements, however, the problem of determining the hydrodynamic behaviour of a single OWSC in the open ocean is still unsolved.

In \S\ref{sec:model} a semi-analytical model for a single OWSC in the open ocean is derived, assuming the fluid inviscid and incompressible, the flow irrotational and the perturbation time-harmonic. The equation of motion of the flap is solved in the frequency domain and the hydrodynamic characteristics of the system are determined. The potential field at large distance from the flap is then analysed, to obtain the asymptotic expression of the radiated and diffracted waves in the far field. Several expressions are derived, which relate the hydrodynamic characteristics to the wave amplitude of the radiated and diffracted waves at infinity. In \S\ref{sec:disc} the model is validated against available numerical results. Then discussion is made on the main parameters of the system, for a layout inspired by the design of a commercial OWSC. In \S\ref{sec:ocvsch} comparison is made between the behaviour of the OWSC in the open ocean and in a channel. This allows to quantify the degree of approximation of experimental models in wave tank, aiming at reproducing the performance of the converter in the open ocean. Finally, in \S\ref{sec:paran} parametric analysis of the OWSC in the open ocean is undertaken to investigate the influence of the flap width on the performance of the device.
\section{Mathematical model}
\label{sec:model}
\subsection{Governing equations}
\begin{figure}
  \centerline{\includegraphics[width=13cm, trim=4cm 11cm 4.5cm 10.5cm, clip]{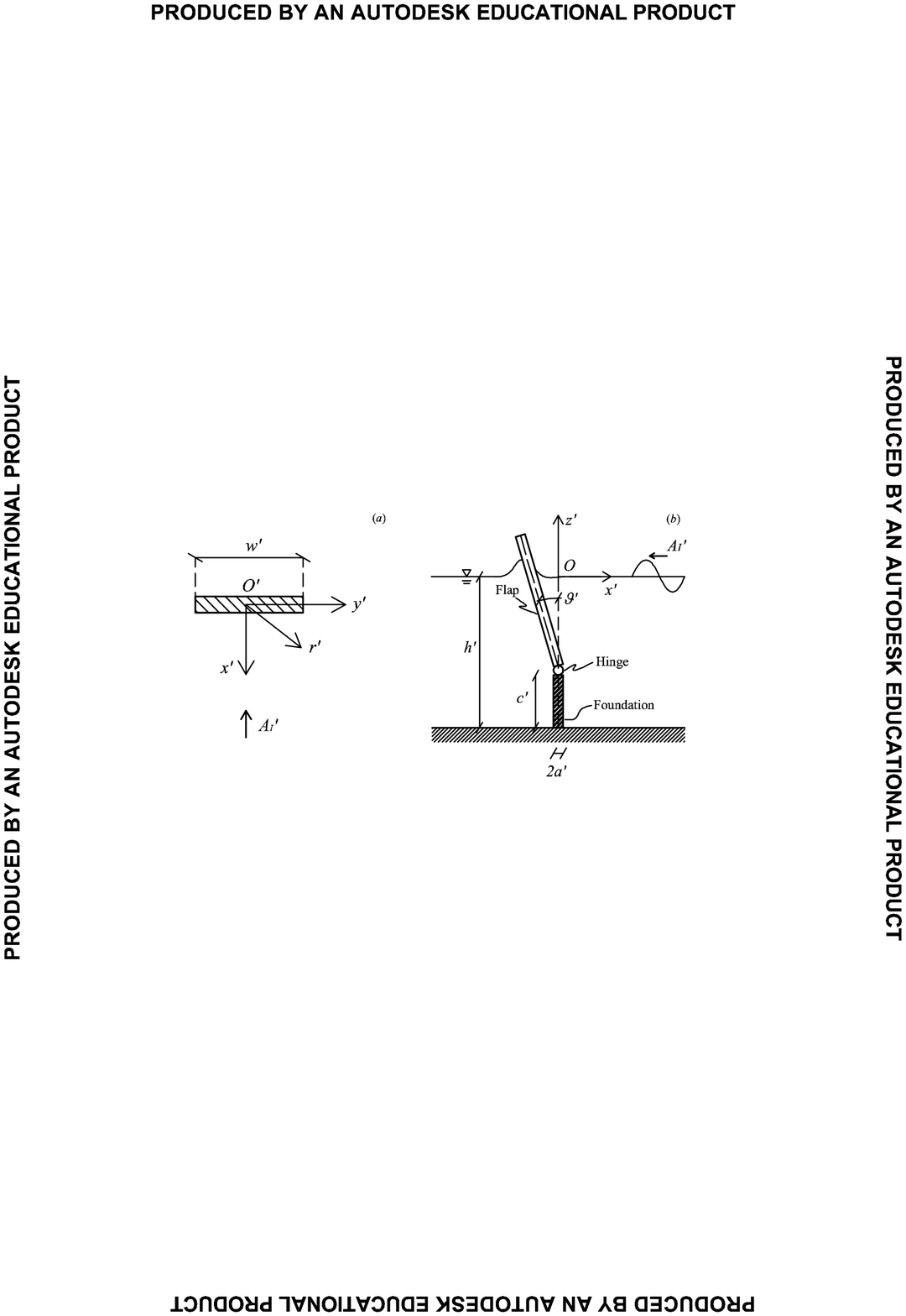}}
  \caption{Geometry of the system in physical variables; ($a$) plan view, ($b$) section.}
\label{fig:geom}
\end{figure}

Referring to figure \ref{fig:geom}$(a,b)$, consider an OWSC in an open ocean of constant depth $h'$; primes denote physical quantities.
The OWSC is modelled as a rectangular flap of width $w'$ and thickness $2a'$. Under the action of monochromatic incident waves of amplitude $A_I'$ and period $T'$, the flap oscillates about a hinge placed upon a bottom foundation of height $c'$. Since the device is designed to work in the nearshore, where the incident wave fronts are almost parallel to the coastline, normal incidence is also assumed. Let $\theta'(t')$ be the pitching amplitude of the device, positive if anticlockwise and let $t'$ denote time. Consider a Cartesian system of reference $O'(x',y',z')$, with the $x'$-axis opposite to the direction of propagation of the incoming waves, the $y'$ axis along the width of the device and the $z'$ axis pointing upwards from the still water level. The origin $O'$ is in the middle of the device at the still water level (see again figure \ref{fig:geom}$a,b$). Let us assume that the fluid is inviscid and incompressible and the flow is irrotational. Hence there exists a velocity potential $\Phi'(x',y',z',t')$ which satisfies the Laplace equation
\begin{equation}
\nabla{'}^2\Phi'(x',y',z',t')=0
\label{eq:lapl}
\end{equation}
in the fluid domain; $\nabla'f'=\left(f'_{,x'},f'_{,y'},f'_{,z'}\right)$ and subscripts with commas denote differentiation with respect to the relevant variable. Let us also assume that the incident wave amplitude is small as compared to the width of the device, which is the reference length scale of the problem: $A'_I/w'\ll 1$. As a consequence, the angular rotation of the flap induced by the incoming waves is small and the behaviour of the system is linear \cite{RD12a}. On the free surface, the linearised kinematic-dynamic boundary condition reads
\begin{equation}
\Phi'_{,t't'}+g\Phi'_{,z'}=0, \quad z'=0,
\label{eq:freesurf}
\end{equation}
with $g$ the acceleration due to gravity. Also, absence of flux at the bottom of the ocean requires
\begin{equation}
\Phi'_{,z'}=0, \quad z'=-h',
\label{eq:bottom}
\end{equation}
while absence of normal flow through the flap yields
\begin{equation}
\Phi'_{,x'}=-\theta'_{,t'}(t') (z'+h'-c')\,H(z'+h'-c'), \quad x'=\pm 0, |y'|<w'/2,
\label{eq:body}
\end{equation}
where the thin-body approximation has been applied \cite[see]{RD12a,LM01}. In expression (\ref{eq:body}), the Heaviside step function $H$ is used to model the absence of flow through the bottom foundation. Finally, the wave field generated by the interaction between the incoming waves and the flap must be outgoing at large distance $r'=\sqrt{x{'}^2+y{'}^2}$ from the body. Let us now introduce the following non-dimensional variables
\begin{equation}
(x,y,z,r)=(x',y',z',r')/w', \, t=\sqrt{g/w'}t',\, \Phi=(\sqrt{gw'}A')^{-1}\Phi', \, \theta=(w'/A')\theta'
\label{eq:vartran}
\end{equation}
and constants 
\begin{equation}
(h,c)=(h',c')/w', \quad A_I=A_I'/A',
\label{eq:constran}
\end{equation}
where $A'=1\,\mathrm{m}$ is the amplitude scale of the incident wave. By making use of (\ref{eq:vartran}) and (\ref{eq:constran}), the governing equation (\ref{eq:lapl}) and the associated boundary conditions (\ref{eq:freesurf})--(\ref{eq:body}) become, respectively,
\begin{equation}
\nabla^2\Phi=0
\label{eq:laplnd}
\end{equation}
in the fluid domain, and
\begin{equation}
\Phi_{,tt}+\Phi_{,z}=0,\quad z=0,
\label{eq:freesurfnd}
\end{equation}
\begin{equation}
\Phi_{,z}=0,\quad z=-h,
\label{eq:bottomnd}
\end{equation}
and finally
\begin{equation}
\Phi_{,x}=-\theta_{,t}(t) (z+h-c)\,H(z+h-c),\quad x=\pm 0, |y|<1/2.
\label{eq:flapnd}
\end{equation}
\subsection{Solution}
Let us assume that the flap performs sinusoidal oscillations of frequency $\omega=\omega'\sqrt{w'/g}=2\pi/T$, for which the time dependence of the system variables can be separated as
\begin{equation}
\theta(t)=\Re\left\lbrace\Theta\,e^{-i\omega t}\right\rbrace, \quad \Phi(x,y,z,t)=\Re\left\lbrace (\phi^I+\phi^D+\phi^R) e^{-i\omega t}\right\rbrace.
\label{eq:timesep}
\end{equation}
In the latter,
\begin{equation}
\phi^I=-\frac{i A_I}{\omega}\frac{\cosh k(z+h)}{\cosh kh}e^{-ikx}
\label{eq:incwav}
\end{equation}
is the spatial potential of the plane incident wave \cite{ME05}, while $\phi^D(x,y,z)$ and $\phi^R(x,y,z)$ are, respectively, the potential of the diffracted and the radiated waves. In (\ref{eq:incwav}), $k$ is the wavenumber, solution of the well-known dispersion relationship $\omega^2=k \tanh kh$. By substituting the decomposition (\ref{eq:timesep}) for the velocity potential into the governing equations (\ref{eq:laplnd})--(\ref{eq:flapnd}), the latter become, respectively,
\begin{equation}
\nabla^2\phi^{(R,D)}=0
\label{eq:laplphird}
\end{equation}
in the fluid domain, the shorthand notation $\phi^{(R,D)}$ denoting either potential, and
\begin{equation}
\phi^{(R,D)}_{,z}-\omega^2\phi^{(R,D)}=0,\quad z=0,
\end{equation}
\begin{equation}
\phi^{(R,D)}_{,z}=0,\quad z=-h,
\end{equation}
and finally
\begin{equation}
\left\lbrace
\begin{array}{c}
\phi^{R}_{,x}\\
\phi^{D}_{,x}
\end{array} \right\rbrace= \left\lbrace
\begin{array}{c}
V (z+h-c) H(z+h-c)\\
-\phi^I_{,x}
\end{array}\right\rbrace, \quad x=\pm 0,\, |y|<1/2.
\label{eq:bcphird}
\end{equation}
In the latter, $V=i\omega\Theta$ is the angular velocity of the flap, depending on the complex amplitude of rotation $\Theta$ (see \ref{eq:timesep}), still unknown. The expressions in (\ref{eq:bcphird}) are motivated by the fact that $\phi^{R}$ is the solution of the problem in the absence of incoming waves, while $\phi^{D}$ is the solution of the problem where the flap is held fixed in incoming waves \cite[see]{ME05}. Finally, $\phi^{(R,D)}$ must be outgoing as $r\rightarrow \infty$. Following the procedure of \cite{RD12a}, let us factor out the vertical dependence of the governing system (\ref{eq:laplphird})--(\ref{eq:bcphird}) by setting
\begin{equation}
\phi^{(R,D)}(x,y,z)=\sum_{n=0}^{\infty} \varphi^{(R,D)}_n(x,y) Z_n(z),
\label{eq:decomp}
\end{equation} 
where
\begin{equation}
Z_n(z)=\frac{\sqrt{2} \cosh \kappa_n(z+h)}{\left(h+\omega^{-2}\sinh^2 \kappa_n h\right)^{1/2}},\quad n=0,1,\dots
\label{eq:Zn}
\end{equation}
are the vertical eigenmodes  of the flat-bottom water wave problem \cite[see]{ME05}, orthogonal along the water column:
\begin{equation}
\int_{-h}^0 Z_n(z) Z_m(z)\, dz=\delta_{nm},\quad n,m \in\mathbb{N},
\label{eq:orth}
\end{equation}
being $\delta_{nm}$ the Kronecker symbol. In (\ref{eq:Zn}), $\kappa_0=k$, while $\kappa_n=i k_n$ are the solutions of the dispersion relation
\begin{equation}
\omega^2=-k_n\tan k_nh, \quad n=1,2,\dots
\end{equation} 
By substituting the decomposition (\ref{eq:decomp}) into (\ref{eq:laplphird}) and (\ref{eq:bcphird}) and using the orthogonality relation (\ref{eq:orth}), we obtain the Helmholtz equation
\begin{equation}
\left( \nabla^2+\kappa_n^2\right)\varphi^{(R,D)}_n=0
\label{eq:helm}
\end{equation}
and the boundary conditions
\begin{equation}
\left\lbrace
\begin{array}{c}
\varphi^R_{n,x}\\
\varphi^D_{n,x}
\end{array}\right\rbrace = \left\lbrace
\begin{array}{c}
Vf_n\\
A_I d_n
\end{array}\right\rbrace, \quad
x=\pm 0, |y|<1/2,
\label{eq:bcvarphi}
\end{equation}
for the unknown two-dimensional velocity potentials $\varphi^{(R,D)}_n$, where $n=0,1,\dots$ is omitted from now on, for the sake of brevity. In expression (\ref{eq:bcvarphi}),
\begin{equation}
f_n=\frac{\sqrt{2}\left[\kappa_n (h-c)\sinh \kappa_n h+\cosh \kappa_n c- \cosh\kappa_n h\right]}{\kappa_n^2 \left(h+\omega^{-2}\sinh^2 \kappa_n h\right)^{1/2}}
\label{eq:fn}
\end{equation} 
and
\begin{equation}
d_n=\frac{k\left(h+\omega^{-2}\sinh^2 \kappa_n h\right)^{1/2}}{\sqrt{2}\omega \cosh \kappa_n h}\,\delta_{0n}
\label{eq:dn}
\end{equation}
are real constants. Finally, the $\varphi^{(R,D)}_n$ must be outgoing disturbances as $r\rightarrow\infty$. The boundary-value problem (\ref{eq:helm})--(\ref{eq:bcvarphi}) set for $\varphi^{(R,D)}_n$ is similar in form to that obtained by Renzi and Dias \cite{RD12a} for an OWSC in a straight channel. In the latter, however, waves are reflected by the lateral walls and can propagate only along the channel axis. Following the general method of \cite{RD12a}, the solution to (\ref{eq:helm})--(\ref{eq:bcvarphi}) is found in \ref{sec:appA} by applying an integral-equation technique based on the Green's theorem. Using an appropriate decomposition of the 3D Green function into a singular part and an analytic remainder allows to solve the plane boundary-value problem for $\varphi_n^{(R,D)}$ and to express the spatial potentials $\phi^{(R,D)}$ (\ref{eq:decomp}) in a semi-analytical form (see \ref{sec:appA} for details).  The radiation potential is given by
\begin{eqnarray}
\phi^R(x,y,z)&=&\frac{-iV}{8}\sum_{n=0}^{\infty}\kappa_n x Z_n(z)\sum_{p=0}^P \alpha_{(2p)n} \nonumber \\
&\times&\int_{-1}^1(1-u^2)^{1/2}U_{2p}(u)\frac{H_1^{(1)}\left(\kappa_n \sqrt{x^2+(y-u/2)^2}\right)}{\sqrt{x^2+(y-u/2)^2}}\,du,
\label{eq:phir}
\end{eqnarray}
where $H_{n}^{(1)}$ is the outgoing Hankel function of first kind and order $n$ and the $U_{2p}$ denote the Chebyshev polynomial of the second kind and even order $2p$, $p=0,1,\dots, P \in\mathbb{N}$. Finally, the $\alpha_{(2p)0}$ are the complex solutions of the system of equations (\ref{eq:lineqrad}), ensuring that the first equation of (\ref{eq:bcphird}) is satisfied. The $\alpha_{(2p)n}$ are obtained with a numerical collocation scheme, therefore the solution (\ref{eq:phir}) is semi-analytical. Analogously, the spatial diffraction potential is given by
\begin{eqnarray}
\phi^D(x,y,z)&=&\frac{-iA_I}{8}\, kx Z_0(z)\sum_{p=0}^P \beta_{(2p)0} \nonumber \\
&\times&\int_{-1}^1(1-u^2)^{1/2}U_{2p}(u)\frac{H_1^{(1)}\left(k \sqrt{x^2+(y-u/2)^2}\right)}{\sqrt{x^2+(y-u/2)^2}}\,du.
\label{eq:phid}
\end{eqnarray}
In the latter, the $\beta_{(2p)n}$ are the complex solutions of the system of equations (\ref{eq:lineqrad}), ensuring that the second expression of (\ref{eq:bcphird}) is satisfied. Let us now analyse the equation of motion of the flap.
\subsection{Flap motion}
In expression (\ref{eq:phir}) the angular velocity of the flap, $V=i\omega\Theta$, is still unknown and must be determined by studying the motion of the body.
The equation of motion of the flap in the frequency domain is that of a damped harmonic oscillator \cite{ME05}, i.e.
\begin{equation}
\left( -\omega^2 I + C-i\omega \nu_{pto}\right)\Theta=\mathcal{F},
\label{eq:mot}
\end{equation}
where $\omega$ is the frequency of oscillation, $I=I'/(\rho w{'}^5)$ is the moment of inertia of the flap, $C=C'/(\rho g w'^4)$ is the flap buoyancy torque and $\nu_{pto}=\nu_{pto}'/(\rho w'^5\sqrt{g/w'})$ is the power take-off (PTO) coefficient; $\rho$ is water density. In expression (\ref{eq:mot}), 
\begin{equation}
\mathcal{F}=i\omega\int_{-h+c}^0\int_{-1/2}^{1/2}\Delta\phi(y,z) (z+h-c)\,dy dz
\label{eq:extor}
\end{equation}
is the complex hydrodynamic torque on the flap, where
\begin{equation}
\Delta\phi=\left[ \phi^D(-0,y,z)+\phi^R(-0,y,z)\right]-\left[\phi^D(+0,y,z)+\phi^R(+0,y,z)\right]
\label{eq:potjump}
\end{equation}
is the jump in potential between the two sides of the flap (see \ref{sec:appA}). By using (\ref{eq:potjump}), (\ref{eq:PnQn}) and (\ref{eq:PnCheb}) in (\ref{eq:extor}), substituting the latter in (\ref{eq:mot}) and performing some algebra, the equation of motion of the flap reads
\begin{equation}
\left[-\omega^2 (I+\mu) +C-i\omega(\nu+\nu_{pto})\right]\Theta= F,
\label{eq:mot2}
\end{equation} 
where
\begin{equation}
F=\frac{-i\omega\pi}{4}A_If_0\beta_{00}
\label{eq:F}
\end{equation}
is the exciting torque,
\begin{equation}
\mu=\frac{\pi}{4}\Re\left\lbrace\sum_{n=0}^\infty f_n\alpha_{0n}\right\rbrace
\label{eq:mu}
\end{equation}
is the added torque due to inertia and finally
\begin{equation}
\nu=\frac{\omega \pi}{4}f_0\Im\left\lbrace \alpha_{00}\right\rbrace
\label{eq:nu}
\end{equation}
is the radiation damping. In the latter, only the propagating mode $n=0$ is present, because higher order eigenmodes correspond to evanescent waves that do not radiate from the flap (see \S\ref{sec:asymp}). In the following, approximated expressions of the radiation and diffraction potentials will be obtained at large distance from the flap.
\subsection{Behaviour in the far field}
\label{sec:asymp}
\begin{figure}
  \centerline{\includegraphics[width=9cm, trim=6cm 11cm 6cm 0cm]{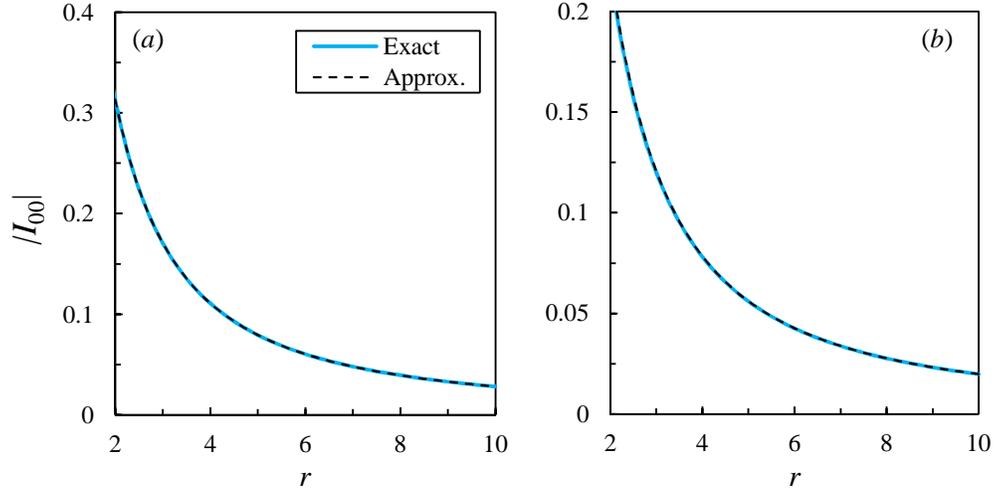}}
  \caption{Approximation of the integral $I_{00}$ (\ref{eq:I0p}) (solid blue line) with expression (\ref{eq:I0papp}) (dashed black line) for large $r$. Parameters are: $(a)$ $k=2$, $(b)$ $k=4$. The geometry of the system is detailed in table \ref{tab:1}.}
\label{fig:approx}
\end{figure}

First, let us consider the radiation potential (\ref{eq:phir}) and rewrite it in polar coordinates: $\phi^R(x,y,z)=\phi^R(r,\gamma,z)$, where 
\begin{equation}
(x,y)=r(\cos\gamma,\sin\gamma)
\label{eq:polcor}
\end{equation}
(see figure \ref{fig:geom}$a$). For an observer far away from the OWSC, the flap behaves as a dipole-mode radiator, so that
\begin{equation}
\phi^R(r,\gamma,z)\simeq\phi^R(r,0,z)\cos\gamma
\label{eq:dipol}
\end{equation}
as $r\rightarrow\infty$. Now perform the change of variable (\ref{eq:polcor}) into (\ref{eq:phir}) and consider the inner integral of $\phi^R(r,\gamma=0,z)$, i.e.
\begin{equation}
I_{np}=\int_{-1}^1 (1-u^2)^{1/2} U_{2p}(u)\frac{H_1^{(1)}\left(\kappa_n r\sqrt{1+\frac{u^2}{4r^2}}\right)}{r\sqrt{1+\frac{u^2}{4r^2}}}\,du.
\label{eq:I0p}
\end{equation}
For $n>0$, then $\kappa_n=i k_n$ and the Hankel function inside the integral (\ref{eq:I0p}) transforms into the modified Bessel function of the second kind $K_1$ \cite[see \S 8.407]{GR07}, which decays exponentially for large $r$. Therefore the terms $n>0$ do not give any significant contribution to the radiated wave field (\ref{eq:dipol}) at large distance from the flap. Physically, such terms correspond to the evanescent modes of the system, which are trapped near the source \cite[see for example]{ME05}. Now consider $n=0$. At large $r$ the Hankel function in $I_{0p}$ (\ref{eq:I0p}) can be approximated as \cite[see]{LM01}
\begin{eqnarray}
H_1^{(1)}\left( kr\sqrt{1+\frac{u^2}{4r^2}}\right)\simeq\sqrt{\frac{2}{\pi kr\sqrt{1+\frac{u^2}{4r^2}} }} \exp\left(ikr \sqrt{1+\frac{u^2}{4r^2}}-i3\pi/4\right).
\label{eq:hankapp}
\end{eqnarray}
Note that the argument of the square roots in the latter expression is $1+O(r^{-2})$. As $r\rightarrow\infty$, the terms $O(r^{-2})$ in (\ref{eq:hankapp}) and in (\ref{eq:I0p}) can be neglected, thus yielding the sought approximated expression
\begin{eqnarray}
I_{0p}&\simeq &\frac{1}{r}\sqrt{\frac{2}{\pi kr}}\,e^{ikr}e^{-i3\pi/4}\int_{-1}^1(1-u^2)^{1/2}U_{2p}(u)\,du \nonumber\\
&=&\frac{\pi}{2}\frac{1}{r}\sqrt{\frac{2}{\pi kr}}\,e^{ikr}e^{-i3\pi/4} \delta_{p0},
\label{eq:I0papp}
\end{eqnarray}
as $r\rightarrow\infty$, where the orthogonality relation of \cite[\S7.343]{GR07} has also been used. Numerical investigation for a typical geometry of the system reveals that (\ref{eq:I0papp}) converges to the exact value (\ref{eq:I0p}), as shown in figure \ref{fig:approx}$(a,b)$.
Now, expressing (\ref{eq:phir}) in polar coordinates (\ref{eq:polcor}), applying the dipole relation (\ref{eq:dipol}) and using the approximated expression (\ref{eq:I0papp}) for the inner integral, yields the expression of the radiation potential in the far field
\begin{equation}
\phi^R(r,\gamma)\simeq \frac{-iV}{\omega}\mathcal{A}^R(\gamma)\,\frac{\cosh k(z+h)}{\cosh kh}\sqrt{\frac{2}{\pi kr}} e^{i(kr-\pi/4)},
\label{eq:phirapp}
\end{equation}
as $r\rightarrow\infty$. In the latter expression,
\begin{equation}
\mathcal{A}^R(\gamma)=\frac{-i \pi}{8\sqrt{2}}\,k\cos\gamma\,\alpha_{00}\frac{\omega\cosh kh}{\left(h+\omega^{-2}\sinh^2 kh\right)^{1/2}}
\label{eq:Ar}
\end{equation}
is the angular variation of the radially spreading wave \cite[see]{ME05}, for unit rotational velocity of the flap. Analogous reasoning applied to the diffraction potential $\phi^D$ (\ref{eq:phid}) yields
\begin{equation}
\phi^D(r,\gamma)\simeq \frac{-iA_I}{\omega}\mathcal{A}^D(\gamma)\,\frac{\cosh k(z+h)}{\cosh kh}\sqrt{\frac{2}{\pi kr}}\,e^{i(kr-\pi/4)}
\label{eq:phidapp}
\end{equation}
at large distance from the plate, where
\begin{equation}
\mathcal{A}^D(\gamma)=\frac{-i \pi}{8\sqrt{2}}\,k\cos\gamma\,\beta_{00}\frac{\omega\cosh kh}{\left(h+\omega^{-2}\sinh^2 kh\right)^{1/2}}
\label{eq:Ad}
\end{equation}
is the angular variation of the diffracted wave, for unit amplitude of the incident wave. In the following section we shall derive some useful relations for the hydrodynamic coefficients of the system.
\subsection{Derivation of relations for the OWSC in the open ocean}
In this section we obtain some relations between the hydrodynamic coefficients of the system, based on the asymptotic analysis of \S\ref{sec:asymp}. Some of these relations are well-known in the literature for 3D bodies of general shape, while some others are new and provide insightful information on the hydrodynamics of the OWSC.
\subsubsection{Relation between $F$ and $\mathcal{A}^R$ (3D Haskind relation)}
Consider the exciting torque $F$ (\ref{eq:F}) and rewrite it as
\begin{equation}
F=\frac{-i\omega \pi}{4}A_I\alpha_{00} d_0
\label{eq:hask1}
\end{equation}
by virtue of (\ref{eq:relalbe}), where $d_0$ is still given by (\ref{eq:dn}). Then isolating $\alpha_{00}$ from (\ref{eq:Ar}) with $\gamma=0$, substituting it into (\ref{eq:hask1}) and performing some simple algebra yields
\begin{equation}
F=\frac{4}{k}A_IC_g\mathcal{A}^R(0),
\label{eq:hask}
\end{equation}
which is the well-known Haskind relation for a 3D floating body \cite[see for example]{ME05}. The latter relates the exciting torque to the wave amplitude radiated by the body in the direction opposite to the incident wave.
\subsubsection{Relation between $F$ and $\mathcal{A}^D$}
Consider again the exciting torque $F$ (\ref{eq:F}). Now isolate the term $\beta_{00}$ from (\ref{eq:Ad}), substitute it into (\ref{eq:F}) and perform the straightforward algebra to obtain
\begin{equation}
F=\frac{4}{k}A_I\mathcal{A}^D(0)\frac{\tanh kh}{k}\left(h-c+\frac{\cosh kc-\cosh kh}{k \sinh kh}\right).
\label{eq:relFAd}
\end{equation}
The latter shows that the exciting torque is also proportional to the amplitude of the wave scattered in the opposite direction to that of the incident wave.
\subsubsection{Relation between $\nu$ and $\mathcal{A}^R$}
Consider now the radiation damping coefficient $\nu$ (\ref{eq:nu}). By isolating $\alpha_{00}$ in (\ref{eq:Ar}) with $\gamma=0$ and substituting it into  (\ref{eq:nu}), the latter gives
\begin{equation}
\nu=\frac{4}{k}\Re\left\lbrace \mathcal{A}^R(0)\right\rbrace \frac{\tanh kh}{k}\left(h-c+\frac{\cosh kc-\cosh kh}{k \sinh kh}\right),
\label{eq:relnuAr}
\end{equation}
which relates the radiation damping to the amplitude of the radiated wave in the forward $x$ direction.
\subsubsection{Relation between $\nu$ and $F$}
Finally let us inspect expression (\ref{eq:relnuAr}). By taking the real part of the Haskind relation (\ref{eq:hask}), isolate $\Re\left\lbrace \mathcal{A}^R(0)\right\rbrace$ and substitute it into (\ref{eq:relnuAr}), so that the latter becomes
\begin{equation}
\nu=\frac{\tanh kh}{k C_g}\left(h-c+\frac{\cosh kc-\cosh kh}{k \sinh kh}\right) \Re\left\lbrace\frac{F}{A_I}\right\rbrace,
\label{eq:relnuF}
\end{equation}
which can be regarded as a particular form of the general Haskind relation (\ref{eq:hask}) tailored to the OWSC. 
Expressions (\ref{eq:hask})--(\ref{eq:relnuF}) have been used to check the numerical evaluations of \ref{sec:appA}. Given the general equation $\mathrm{l.h.s}=\mathrm{r.h.s}$, the relative error
$$\epsilon=\frac{|\mathrm{l.h.s.}-\mathrm{r.h.s.}|}{|\mathrm{r.h.s.}|}$$
is defined to assess the accuracy of computations. For a typical OWSC configuration (see table \ref{tab:1}), taking $\mathrm{max}\,n=2$ and $\mathrm{max}\,p=5$ in (\ref{eq:lineqrad}) allows to obtain a maximum relative error $\epsilon=O(10^{-15})$ for expressions (\ref{eq:hask})--(\ref{eq:relnuF}).
In the following section we shall analyse the performance of the device.
\subsection{Wave energy extraction}
Consider the equation of motion of the flap (\ref{eq:mot2}). The latter describes a damped harmonic oscillator in the frequency domain. As a consequence, the average generated power over a period is
\begin{equation}
P=\frac{|F|^2}{4(\nu_{pto}+\nu)},
\end{equation}
provided the PTO system is designed such that
\begin{equation}
\nu_{pto}=\sqrt{\frac{\left[C-(I+\mu)\omega^2 \right]^2}{\omega^2}+\nu^2},
\label{eq:nures}
\end{equation}
which corresponds to the optimum PTO damping \cite[see for example]{FA02}. To assess the performance of the OWSC, the capture factor
\begin{equation}
C_F=\frac{P}{\frac{1}{2}A_I^2 C_g}=\frac{|F|^2}{2A_I^2 C_g (\nu_{pto}+\nu)}
\label{eq:CF}
\end{equation}
is defined as the ratio between the power output of the device and the  power of the incident wave per unit width of the device, where
\begin{equation}
C_g=\frac{\omega}{2k}\left(1+\frac{2kh}{\sinh 2kh}\right)
\end{equation}
is the group velocity of the incident wave in non-dimensional variables. Furthermore, if the flap is designed such that its buoyancy torque $C$ and inertia torque $I$ satisfy the relation
\begin{equation}
\omega^2 (I+\mu)=C,
\label{eq:res}
\end{equation} 
i.e.  the flap is tuned to resonance with the incident waves, $\nu_{pto}=\nu$ from (\ref{eq:nures}) and the capture factor (\ref{eq:CF}) attains its maximum value
\begin{equation}
C_F^{max}=\frac{|F|^2}{4A_I^2 C_g\nu}=\frac{|\mathcal{A}^D(0)|^2}{k \Re\left\lbrace \mathcal{A}^D(0)\right\rbrace},
\label{eq:CFmax}
\end{equation}   
where (\ref{eq:relFAd}) and (\ref{eq:relnuF}) have also been used. Note that the maximum capture factor (\ref{eq:CFmax}) depends on the amplitude of the diffracted wave in the direction opposite to that of the incident wave and can be determined directly by solving the diffraction problem for $\phi^D$ (\ref{eq:phid}). In the following, the mathematical model presented in this section will be validated against available numerical results. Then discussion will be made on the hydrodynamic coefficients of a typical OWSC configuration. Afterwards, the behaviour of a single OWSC in the open ocean will be compared to that of an OWSC in a channel, which reproduces a common layout for experimental studies in wave tank. Finally, parametric analysis will be performed to investigate the influence of the flap width on the performance of the OWSC in the open ocean.
\section{Discussion}
\label{sec:disc}
\subsection{Validation}
\begin{figure}
  \centerline{\includegraphics[width=9cm, trim=6.5cm 11cm 6.5cm -1.5cm]{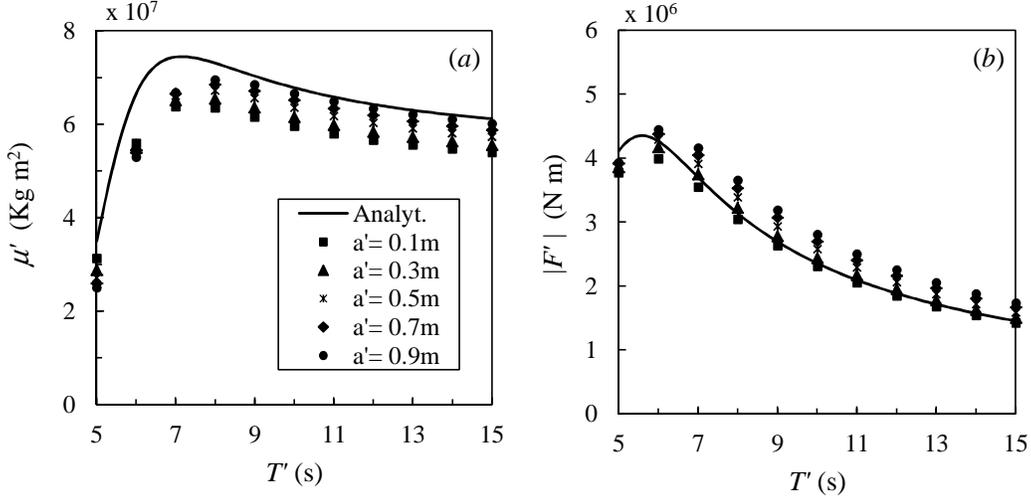}}
  \caption{Comparison of the semi-analytical model of \S\ref{sec:model} (solid lines) with the numerical model of van't Hoff \cite{JO09} (dots), for several thicknesses of the plate $2a'$. ($a$) added inertia torque (\ref{eq:mu}); ($b$) exciting torque (\ref{eq:F}). Values are in physical variables.}
\label{fig:val2}
\end{figure}

In this section we shall compare the results of the mathematical model of \S\ref{sec:model} with available numerical data. The geometry of the system chosen for comparison is reported in table \ref{tab:1}.
\begin{table}\centering
\begin{tabular}{c c}
\hline
Flap width (m) &  Foundation height (m) \\
$w'$ & $c'$\\ 
18 & 1.5 \\ \hline 
Ocean depth (m) & Amplitude incident wave (m) \\
  $h'$ & $A'_I$ \\
 10.9 & 0.3\\ \hline
\end{tabular}
\caption{Geometry of the system and parameters of the incident wave field chosen for comparison with numerical data.}
\label{tab:1}
\end{table}
Such dimensions are inspired by the design of the Oyster 1$^{\mathrm{TM}}$ wave energy converter developed by Aquamarine Power Ltd \cite[APL, see]{APL}. Values of $C$ and $I$ have been obtained by private communication with APL. Note that, since $A_I'/w'\simeq 0.02\ll 1$, the regime of the system is linear. In figure \ref{fig:val2} $(a,b)$ comparison is made between the semi-analytical results of \S\ref{sec:model} and the numerical data of van't Hoff \cite{JO09} for a single OWSC in the open ocean. 
The latter have been obtained with the software package WAMIT \cite{WA} for the linear analysis of fluid-structure interaction. In \cite{JO09} the flap was modelled as a rectangular, $18\,\mathrm{m}$ wide box standing on a triangular prism pointing downwards, hinged on a $1.5\,\mathrm{m}$ tall platform in a $10.9\,\mathrm{m}$ deep ocean. Figures \ref{fig:val2}$(a)$  and \ref{fig:val2}$(b)$ show, respectively, the values of the added inertia torque $\mu'=\mu \rho w'^5$ and the magnitude of the exciting torque $|F'|=|F|\rho g A'w'^3$ against the period of the incident waves for the semi-analytical model of \S\ref{sec:model} and the numerical model of \cite{JO09}.  The agreement between both models is very good for the exciting torque and satisfactory for the added inertia torque. For the latter, minor discrepancies (see figure \ref{fig:val2}$a$) are likely due to the difference in shape between the flaps used in the models. Let us now analyse the performance of the device.
\subsection{Performance analysis}
\begin{figure}
  \centerline{\includegraphics[width=9cm, trim=7cm 12cm 6cm 0cm]{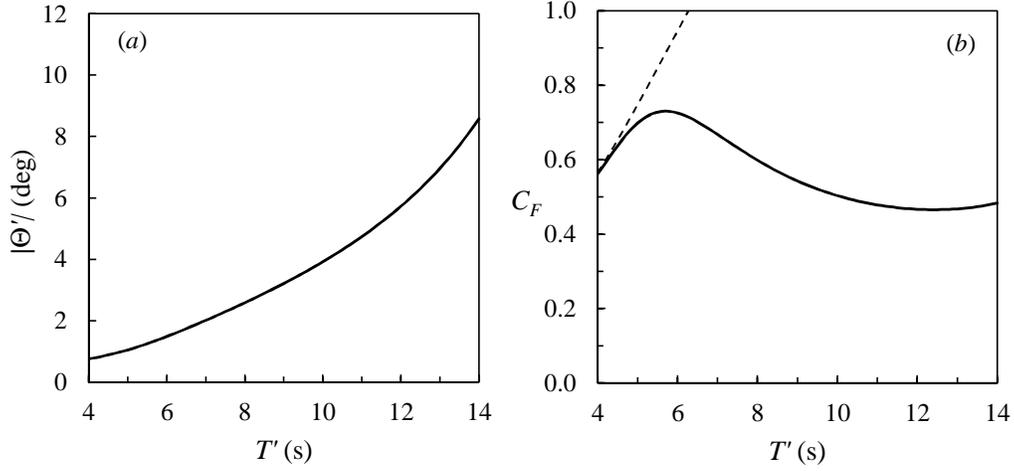}}
  \caption{Behaviour of $(a)$ amplitude of rotation in degrees and $(b)$ capture factor (\ref{eq:CF}) with the period of the incident wave in physical variables, for the geometry of table \ref{tab:1}. In $(b)$ the dashed lines represents the maximum theoretical capture factor $C_F^{max}$ (\ref{eq:CFmax}).}
\label{fig:bodymot}
\end{figure}

Figure \ref{fig:bodymot}($a,b$) shows the magnitude of the rotation $|\Theta'|=(A_I'/w')|\Theta|$ and the capture factor $C_F$ for the configuration of table \ref{tab:1}. 
While the angular displacement of the flap increases monotonically with the incident wave period $T'$, the capture factor instead is larger in short periods. The largest capture factor is $\mathrm{max} \left\lbrace C_F\right\rbrace\simeq 0.73$ at $T'\simeq 5.7 \mathrm{s}$ and is of the same order of magnitude as that predicted by Whittaker and Folley \cite{WF12} for a similar OWSC configuration. For real devices, more complex flap shapes and foundation systems could help to achieve even larger figures \cite{BA12}. Note that the behaviour of the capture factor of figure \ref{fig:bodymot}($b$) resembles very closely that of the exciting torque shown in figure \ref{fig:val2}($b$). Hence the performance of the OWSC is essentially dominated by the exciting torque, as already outlined by \cite{HE10}. In other words, the dynamics of the OWSC is governed by diffractive processes. It is then clear that PA and QPA dynamics, governed by radiative processes \cite{FA02}, are inadequate to describe the behaviour of large flap-type WECs like the OWSC. In figure \ref{fig:bodymot}($b$), the maximum theoretical capture factor $C_F^{max}$ (\ref{eq:CFmax}) is also plotted for comparison with the actual value $C_F$ (\ref{eq:CF}). The two curves are very close for short periods. Then $C_F^{max}$ increases almost linearly in longer waves and becomes much larger than $C_F$. This happens since the OWSC is not designed tuned to resonance, i.e. the flap buoyancy torque $C$ and inertia $I$ do not satisfy the resonance condition (\ref{eq:res}). As a consequence, $\nu_{pto}>\nu$ (see \ref{eq:nures}) and $C_F<C_F^{max}$ (see \ref{eq:CF}). This choice can be justified by considering that, if the resonance condition (\ref{eq:res}) occurred, then the stroke of the flap would exceed by far the amplitude of the incident wave, resulting in a situation not compatible with the power take-off system \cite{WF12,HE10}. In conclusion, $C_F^{max}$ (\ref{eq:CFmax}) can be used in short periods for a quick performance assessment  of the device. Then the analysis must be refined by calculating $C_F$ with the full expression (\ref{eq:CF}). In the following section, we shall compare the behaviour of the OWSC in the open ocean and in a channel. This will allow us to assess the influence of the channel lateral walls on the performance of the system in experimental studies conducted in wave tanks.
\section{OWSC in the open ocean vs. OWSC in a channel}
\label{sec:ocvsch}
\begin{figure}
  \centerline{\includegraphics[width=7.4cm, trim=8cm 11cm 7.5cm 0cm]{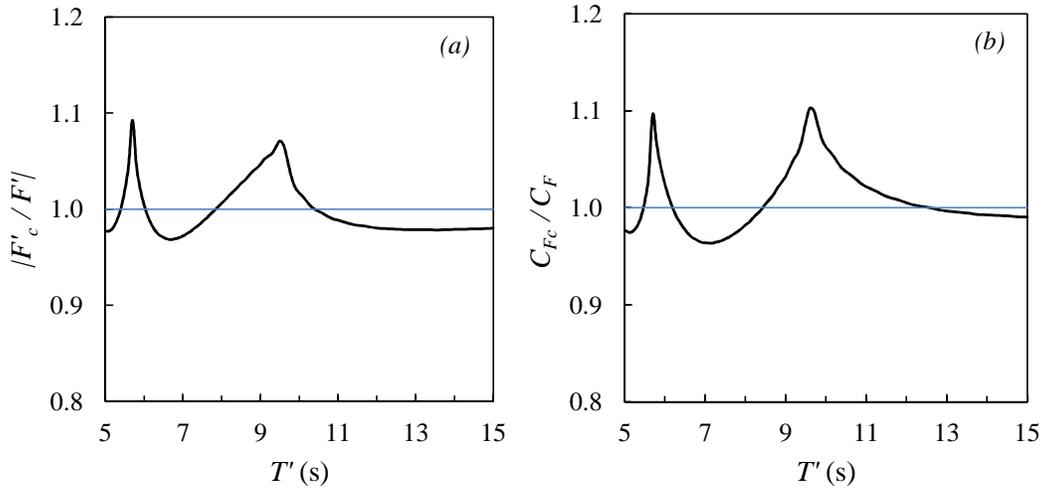}}
  \caption{Behaviour of $(a)$ exciting torque ratio and $(b)$ capture factor ratio between the open ocean model of \S\ref{sec:model} and the channel model of Renzi and Dias \cite{RD12a}, versus the period of the incident wave. The geometry of the system is detailed in table \ref{tab:1}. The blockage ratio in the channel model is $w=0.2$.}
\label{fig:chancomp}
\end{figure}

\begin{figure}
  \centerline{\includegraphics[width=6cm, trim=7.5cm 8cm 7cm -2cm]{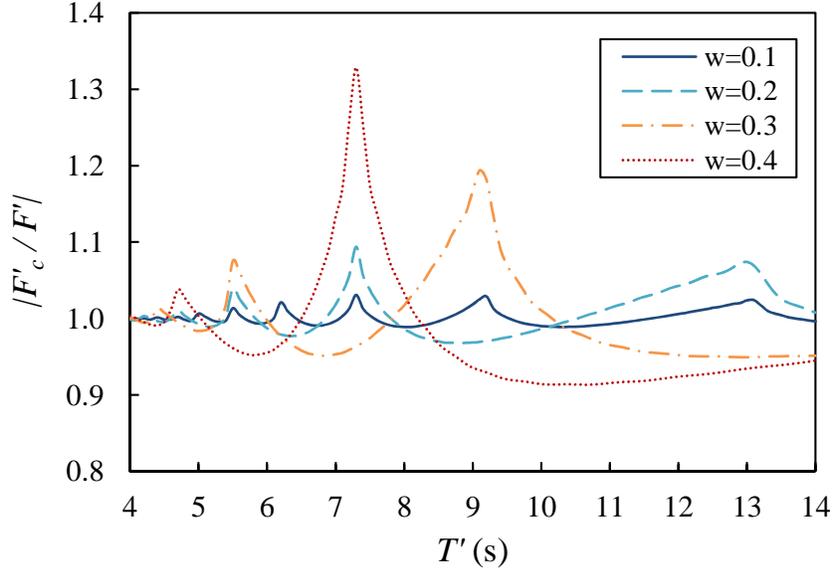}}
  \caption{Behaviour of the exciting torque ratio between the open ocean model of \S\ref{sec:model} and the channel model of Renzi and Dias \cite{RD12a} for several blockage ratios, versus the period of the incident wave. The geometry of the system is detailed in table \ref{tab:1}.}
\label{fig:chanpar}
\end{figure}

Let us compare the behaviour of the OWSC in the open ocean (see \S\ref{sec:model}) versus the behaviour of the same device in a channel of width $b'$ \cite[see]{RD12a}. Such a layout is commonly used in wave tank experiments aiming at reproducing the dynamics of the system in the open ocean \cite[see for example]{WF12, HE10}. The geometry chosen for comparison is again that of table \ref{tab:1}. In the channel model, the channel width is $b'=91.6\,\mathrm{m}$ and the relevant blockage ratio is $w=w'/b'\simeq0.2$. The latter corresponds to that used in wave tank experiments at Queen's University Belfast on a $1:20$ scale Oyster 1 WEC \cite[for details see]{HE10, HE08}. In figure \ref{fig:chancomp}($a$) the ratio $|F'_c/F'|$ between the exciting torque in the channel $|F'_c|$ and in the ocean $|F'|$ is plotted against the period of the incident wave in physical variables. 
The effect of the lateral walls in the channel configuration is revealed by the spikes occurring at the resonant periods of the channel sloshing modes (see table \ref{tab:2}). 
\begin{table}[t]
\begin{center}
\begin{tabular}{c c c c c c}
\hline
$m$  &  1 & 2 & 3 & 4 & 5             \\
\hline
$T_m\,(\mathrm{s})$  & 9.6  & 5.7 & 4.5 &3.8 & 3.4\\
\hline
\end{tabular}
\end{center}
\caption{Resonant periods of the first five sloshing modes for the geometry of table \ref{tab:1} in a channel of width $b'=91.6\,\mathrm{m}.$}
\label{tab:2}
\end{table}
Such periods correspond to the cut-off wavelengths $\lambda'_m=b'/m$, $m=1,2,\dots$, at which the sloshing modes turn from propagating to trapped near the device \cite[for details see]{RD12a}. For the chosen geometry, the influence of the lateral walls results in a $10\%$  increase of the exciting torque in the channel, at the resonant peaks, with respect to the open ocean values (see again figure \ref{fig:chancomp}$a$). Analogous results have been also obtained by Chen \cite{CH94} for the horizontal loads on a truncated vertical cylinder. Note that a similar behaviour occurs also for the ratio $C_{Fc}/C_F$ of the capture factor in the channel $C_{Fc}$ and in the open ocean $C_F$, as depicted in figure \ref{fig:chancomp}($b$). The increase of the capture factor is of the same order of magnitude - about $10\%$ - than the increase of the exciting torque. It is then clear that care should be taken when employing the results obtained in a wave tank to predict the behaviour of the OWSC in the open ocean, as already noted by Renzi and Dias \cite{RD12a}. To quantify the degree of approximation of such practice, let us further investigate the effect of changing the width of the channel in the wave tank model. Figure \ref{fig:chanpar} shows the behaviour of the exciting torque ratio $|F'_c/F'|$ against the period of the incident wave in physical variables, for several blockage ratios. 
When $w$ is small, the channel is much larger than the flap and a number of resonant peaks occur (see curve $w=0.1$ in figure \ref{fig:chanpar}). However, the influence of the lateral walls is not appreciable and $|F'_c/F'|\simeq 1$. When $w$ increases, the number of resonant peaks in the selected period range decreases, while the ratio $|F'_c/F'|$ increases at peaks. The deviation of the latter ratio from unity is still negligible (about $10\%$) for $w=0.2$, but is definitely substantial for $w\geq 0.3$. Therefore we recommend a blockage ratio not larger than $0.2$ in wave tank tests aiming at reproducing the behaviour of the device in the open ocean. In the following section we shall perform a parametric analysis of the OWSC in the open ocean with respect to the flap width $w'$.
\section{Parametric analysis of the OWSC in the open ocean}
\label{sec:paran}
\begin{figure}
  \centerline{\includegraphics[width=8cm, trim=9cm 9cm 8.7cm -2cm]{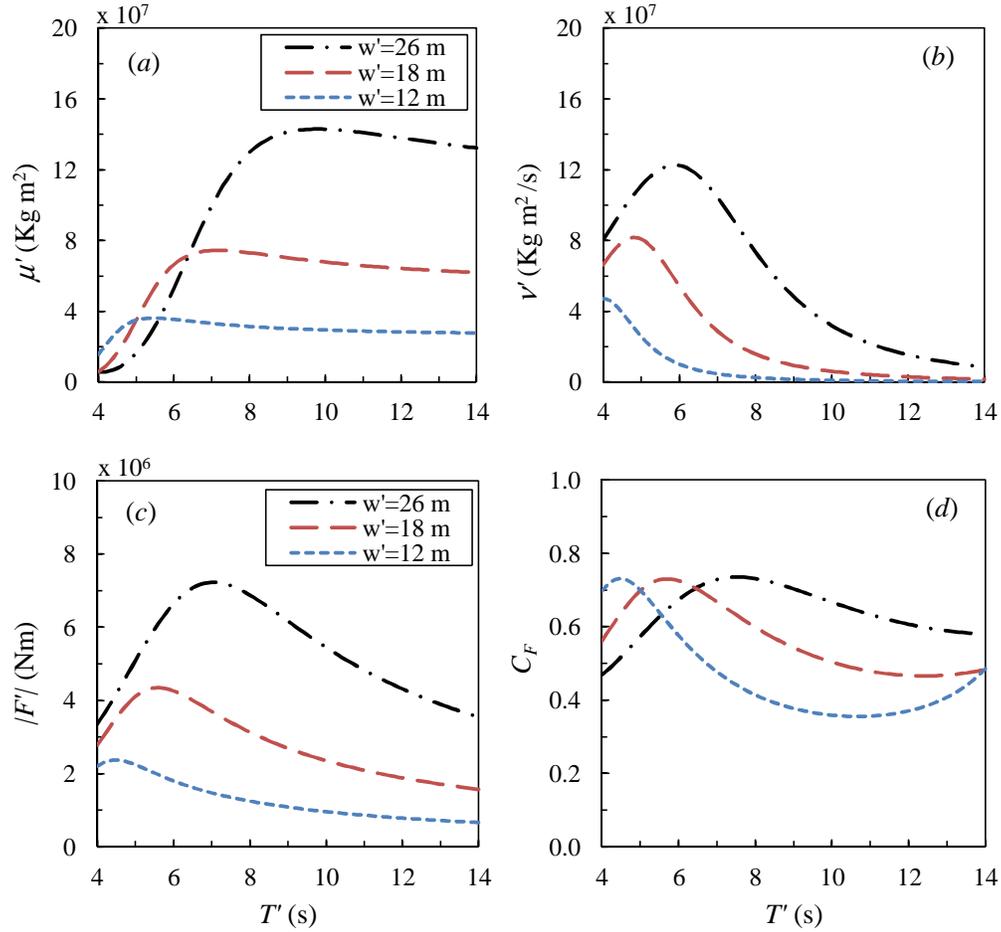}}
  \caption{Behaviour of $(a)$ added inertia torque, $(b)$ radiation damping, $(c)$ magintude of the exciting torque and $(d)$ capture factor versus the period of the incident wave. The geometry of the system is detailed in table \ref{tab:1}. Three different flap widths have been considered.}
\label{fig:wpar}
\end{figure}

Here we undertake a parametric analysis of the system to investigate the influence of the flap width $w'$ on the behaviour of the OWSC. In figure \ref{fig:wpar}($a-d$) the added inertia torque $\mu'$, the radiation damping $\nu'$, the magnitude of the exciting torque $|F'|$ and the capture factor $C_F$ are plotted against the period of the incident waves in physical variables, for three different widths of the device, namely $w'=12, 18, 26 \mathrm{m}$. 
The foundation height, the ocean depth and the incident wave amplitude are still those of table \ref{tab:1}. Generally, increasing the width $w'$ determines all the hydrodynamic characteristics, $\mu'$, $\nu'$ and $|F'|$, to increase. The larger the width of the flap, the larger are the peaks of said parameters, as shown in figure \ref{fig:wpar}($a-c$). A larger flap width, in fact, corresponds to a larger surface area, which in turn determines higher inertia torque, radiative capacity and exciting torque \cite[see]{WF12,HE10}. Concerning the capture factor $C_F$, increasing the width of the flap determines both a migration of the peak values towards larger periods and a broadening of the curve bandwidth, as shown in figure \ref{fig:wpar}($d$). The peak values of the capture factor curve, however, remain almost constant. This behaviour is different from that noticed by Renzi and Dias \cite{RD12a} for an OWSC in a channel. There, the wider flaps have also a larger resonant peak, provided the blockage ratio is still small \cite[otherwise the dynamics is quasi-2D and $C_F\rightarrow 1/2$, see]{RD12a}. This is due to the selective behaviour of the channel sloshing modes, for which the most powerful resonance occurs with large flaps. In the open ocean, on the other hand, the radial spreading of energy to infinity inhibits such resonant mechanisms and prevents the occurrence of resonant peaks in the capture factor curve. Nevertheless, increasing the OWSC width in the open ocean has still the beneficial effects of broadening the bandwidth and increasing the peak period of curve (see again figure \ref{fig:wpar}$d$). In practical applications in random seas, this fact could be exploited by designing the width of the flap so that the resulting capture factor bandwidth couples well to the wave spectrum of the incident sea.

\section{Conclusions}
A semi-analytical potential flow model is derived for a single OWSC in the open ocean. Application of the Green's integral theorem in the fluid domain yields a system of hypersingular integral equations for the jump in potential between the two sides of the flap. The system is solved via a series expansion in terms of the Chebyshev polynomials of the second kind and even order. The hydrodynamic parameters of the system are then fully characterised and the equation of motion of the flap is solved. Asymptotic expressions of the velocity potential are obtained for both the diffracted and radiated wave fields. Several relationships are found between the hydrodynamic parameters and the amplitude of the radiated and diffracted waves in the far field. Some of them are well known in the literature, while others are new and point out the distinguishing features of the OWSC with respect to other WECs. Analysis for a specific OWSC configuration, inspired by the design of the Oyster 1$^{\mathrm{TM}}$ WEC, shows that the capture factor of the system is governed by the exciting torque acting on the flap. High levels of capture factor can be attained, even though the OWSC is not usually tuned to resonance with the incident wave field. The behaviour of the OWSC in the open ocean is then compared with that in a channel. This allows to quantify the degree of accuracy of wave tank experiments aiming at reproducing the behaviour of the device in the open ocean. We found that blockage ratios less than $0.2$ guarantee a good agreement between the open-ocean and the channel models. Finally, parametric analysis of the OWSC in the open ocean reveals that the hydrodynamic characteristics of the system, i.e. exciting torque, added inertia torque and radiation damping, all increase by increasing the width of the flap $w'$. The peak capture factor, instead, shows no appreciable variations with $w'$. However, increasing the flap width has the beneficial effect of enlarging the bandwidth of the capture factor curve. This phenomenon could be exploited in random seas to tune the monochromatic capture factor curve to the incident wave spectrum.  \\

This study was funded by Science Foundation Ireland (SFI) under the research project ``High-end computational modelling for wave energy systems''.
 \appendix

\section{Semi-analytical solution}
\label{sec:appA}
In this section we find the solution to the radiation and scattering problems of \S\ref{sec:model} by using the Green integral theorem in the fluid domain. The method we employ follows the same procedure devised by Renzi and Dias \cite{RD12a} for a flap in a channel. Consider the plane potentials $\varphi_n^{(R,D)}$, solutions of the system (\ref{eq:helm})--(\ref{eq:bcvarphi}) and outgoing as $r\rightarrow\infty$. Application of the Green integral theorem in a large circular region enclosing the flap in the $(x,y)$ plane \cite[see]{ME97} yields
\begin{equation}
\varphi_n^{(R,D)}=\int_{-1/2}^{1/2} \Delta\varphi_n^{(R,D)} (\eta) \left. G_{n,\xi}(x,y;\xi,\eta)\right|_{\xi=0}\, d\eta.
\label{eq:Green}
\end{equation}
In the latter,
\begin{equation}
G_n(x,y;\xi,\eta)=\frac{1}{4i}H_0^{(1)}\left(\kappa_n \sqrt{(x-\xi)^2+(y-\eta)^2}\right)
\label{eq:Gf}
\end{equation}
is the Green function in the $(x,y)$ plane \cite[see]{LM01}, singular at $(x,y)=(\xi,\eta)$ and outgoing as $r\rightarrow\infty$, while
\begin{equation}
\Delta\varphi^{(R,D)}_n(y)=\varphi_n^{(R,D)}(-0,y)-\varphi_n^{(R,D)}(+0,y)
\end{equation}
is the jump in the plane radiation and diffraction potentials across the plate. Following the procedure of \cite{RD12a}, apply the boundary condition (\ref{eq:bcvarphi}) to the spatial potentials (\ref{eq:Green}). Then expand the Hankel function inside $G_n$ (\ref{eq:Gf})  according to \cite[\S 8.444]{GR07}, perform the change of variable $u=2\eta$ and define 
\begin{equation}
(P_n,Q_n)(u)=(\Delta\varphi_n^{R},\Delta\varphi_n^D)(\eta).
\label{eq:PnQn}
\end{equation}
With these manipulations, expression (\ref{eq:bcvarphi}) finally becomes
\begin{eqnarray}
\mathrm{H}\int_{-1}^{1} \left\lbrace\begin{array}{c}
P_n(u)\\Q_n(u)
\end{array}\right\rbrace(v_0-u)^{-2}\,du\nonumber \\
+\left\lbrace\begin{array}{c}\mathcal{K}_n\left[P_n(u)\right]\\
\mathcal{K}_n\left[Q_n(u)\right]
\end{array}\right\rbrace=-\pi w\left\lbrace\begin{array}{c} Vf_n \\ A_0\,d_n\end{array}\right\rbrace, \quad |y|<1/2.
\label{eq:hypersingshort}
\end{eqnarray}
In the latter, which is formally equivalent to (B15) of \cite{RD12a}, the symbol $\mathrm{H}\int$ indicates the Hadamard finite-part integral \cite[see]{RD12a, LM01} and
\begin{equation}
\mathcal{K}_n[f(u)]=\frac{i\pi\kappa_n}{4}\int_{-1}^{1} f(u) \frac{R_n(\frac{k_n}{2} |v_0-u|)}{|v_0-u|}\, du,
\label{eq:Kn}
\end{equation}
where $v_0=v_0(y)=2y\in (-1,1)$ and
\begin{eqnarray}
R_n(\alpha) &=& J_1(\alpha)\left[1+\frac{2i}{\pi}\left(\ln\frac{\alpha}{2}+\gamma\right)\right]\nonumber\\
&-&\frac{i}{\pi}\left[\frac{\alpha}{2}+\sum_{j=2}^{+\infty}\frac{(-1)^{j+1}(\alpha/2)^{2j-1}}{j! (j-1)!}\left(\frac{1}{j}+\sum_{q=1}^{j-1}\frac{2}{q}\right)\right],
\label{eq:Rn}
\end{eqnarray}
 $J_1(x)$ is the Bessel function of first kind and first order and $\gamma=0.577\:215\dots$ the Euler constant. Note that $\mathcal{K}_n$ (\ref{eq:Kn}) has a convergent kernel. Hence expression (\ref{eq:hypersingshort}) has a singularity only in the finite-part integral of the left-hand side. In order to solve (\ref{eq:hypersingshort}) in terms of the jump in potentials $P_n$ and $Q_n$, we seek for solutions of the type
\begin{equation}
\left\lbrace \begin{array}{c}
P_n(u)\\ Q_n(u)
\end{array}\right\rbrace=\left\lbrace\begin{array}{c}V \\A_0\end{array}\right\rbrace\left(1-u^2\right)^{1/2}\sum_{p=0}^{+\infty}\left\lbrace \begin{array}{c} \alpha_{pn} \\ \beta_{pn} \end{array} \right\rbrace U_{2p}(u),
\label{eq:PnCheb}
\end{equation}
where the $\alpha_{pn}$ and $\beta_{pn}$ are unknown complex constants and the $U_{2p}$ are the Chebyshev polynomials of the second kind and even order $2p$, $p\in\mathbb{N}$. Substituting the series expansion (\ref{eq:PnCheb}) into the singular equation (\ref{eq:hypersingshort}) and performing the algebra yields finally
\begin{equation}
\sum_{p=0}^{\infty}\left\lbrace\begin{array}{c}\alpha_{pn}\\\beta_{pn}\end{array}\right\rbrace C_{pn}(v_0)= - \left\lbrace\begin{array}{c} f_n\\d_n\end{array}\right\rbrace,\quad v_0\in(-1,1),
\label{eq:lineqrad}
\end{equation}
with
\begin{eqnarray}
C_{pn}(v_0)&=& -(p+1)\,U_p(v_0)+\frac{i \kappa_n}{4}\int_{-1}^{1} \left(1-u^2\right)^{1/2} U_p(u)\nonumber\\
&\times&\frac{R_n(\frac{\kappa_n}{2}|v_0-u|)}{|v_0-u|}\,du.
\label{eq:Cpn}
\end{eqnarray}
The linear systems of (\ref{eq:lineqrad}) can be now solved numerically by truncating the series to a finite number $P<\infty$ and by taking a finite number of evaluation points $v_0=v_{0j}$, with $j=0,1,\dots P$. A fast numerical convergence of (\ref{eq:lineqrad}) is guaranteed when the $v_{0j}$ are the zeros of the first-kind Chebyshev polynomials \cite[see]{LM01}
\begin{equation}
v_{0j}=\cos\frac{(2j+1)\pi}{2P+2},\quad j=0,1,\dots,P,
\label{eq:voj}
\end{equation}
for which (\ref{eq:lineqrad}) reduces to two $(P+1)\times (P+1)$ truncated algebraic systems for each $n$. The latter can be easily solved to determine the $\alpha_{pn}$ and the $\beta_{pn}$. Finally, some solvability conditions must be satisfied to ensure the uniqueness of the solution to (\ref{eq:lineqrad}). For $n>0$, $d_n=0$ (see \ref{eq:dn}), so that it must be $\beta_{pn}=0$ to ensure uniqueness of the $\alpha_{pn}$. Physically, this means that the diffraction problem does not admit evanescent modes. For $n=0$, instead, the solution to (\ref{eq:lineqrad}) is unique if 
\begin{equation}
\beta_{p0}=\alpha_{p0} d_0/f_0.
\label{eq:relalbe}
\end{equation}
The plane radiation and scattering potentials can now be obtained by substituting the found jumps in potential (equations \ref{eq:PnQn} and \ref{eq:PnCheb}) into the Green's theorem (\ref{eq:Green}) and by summing up all the vertical eigenmodes according to (\ref{eq:decomp}).





\end{document}